\documentclass{mscs}

\usepackage{amssymb}
\usepackage{amsfonts}
\usepackage{amsmath}
\usepackage{graphicx}

\newtheorem{example}{Example}[section]

\newtheorem{proposition}[example]{Proposition}

\def\LU{{\mbox{\tiny LUT}}}
\def\LUT{{\mbox{\tiny LUT}}}
\def\LSU{{\mbox{\tiny LSUT}}}
\def\LSUT{{\mbox{\tiny LSUT}}}
\def\SLOCC{{\mbox{\tiny SLOCC}}}
\def\Det{{\bf Det}}

\def\<{\langle}
\def\>{\rangle}
\def\H{{\cal H}}

\def\C{{\mathbb C\, }}

\def\uu{{\bf u}}

\def\SG{{\mathfrak S}}
\def\PP{{\mathbb P}}
\def\xx{{\bf x}}

\def\gg{{\bf g}}
\def\aa{{\bf a}}
\def\tr{{\rm tr\,}}
\def\Cov{{\rm Cov}}
\def\Inv{{\rm Inv}}

\begin{document}

\title[Unitary invariants of  qubit systems]{Unitary invariants of qubit systems}

\author[J.-G. Luque, J.Y. Thibon and F. Toumazet]
{{\sc J\ls e\ls a\ls n\ls -\ls G\ls a\ls b\ls r\ls i\ls e\ls l\ns L\ls u\ls q\ls u\ls e$^1$,
 J\ls e\ls a\ls n\ls -\ls Y\ls v\ls e\ls s\ns T\ls h\ls i\ls b\ls o\ls n$^1$} and
 {\sc F\ls r\ls\'e\ls d\ls\'e\ls r\ls i\ls c\ns T\ls o\ls u\ls m\ls a\ls z\ls e\ls t$^2$}\\
$^1$Institut Gaspard Monge, Universit\'e de
Marne-la-Vall\'ee,\addressbreak F-77454 Marne-la-Vall\'ee cedex, France.\addressbreak
email: \{luque,jyt\}@univ-mlv.fr
\addressbreak $^2$Laboratoire
d'Informatique de Paris Nord, Institut
Galil\'ee-Universit\'e
 Paris 13/Paris Nord,\addressbreak 99 av. J.-B. Clement 93430 Villetaneuse,
France.\addressbreak email: ft@lipn.univ-paris13.fr}

\maketitle
\date{}
\begin{abstract}
We give an algorithm allowing to construct bases of  
local unitary invariants of pure $k$-qubit states
from the knowledge of 
polynomial covariants  of the group of invertible local filtering operations.
The simplest invariants obtained in this way  are explicited and compared to
various known entanglement measures. Complete sets of generators are obtained
for up to four qubits, and the structure of the invariant algebras is
discussed in detail.
\end{abstract}

\section{\label{intro}Introduction}

From a mathematical point of view,  Quantum Information Theory deals with finite
dimensional Hilbert spaces, the state spaces of finite $k$-partite
systems, which have the special form
\begin{equation} 
\H=V_1\otimes V_2\otimes\cdots\otimes V_k\,,
\end{equation}
where $V_i$ is the finite dimensional state space of the $i$th part
(or particle) of the system, most of the time assumed to be a
{\em qubit}, which means that $\dim\, V_i=2$.

The interesting non-classical behaviors on which the theory is based
already occur for two-qubit systems, for the so-called
{\em entangled states}, those $\psi\in V_1\otimes V_2$ which cannot
be written in the form $v_1\otimes v_2$.
The properties of such states 
are the basis of the EPR paradox \cite{EPR}, and since its discovery,
the entanglement phenomenon has been thoroughly investigated by  physicists,
cf. \cite{Bel,HSH,AGR,BW},  and more recently by mathematicians, 
e.g., \cite{Bry, Kly, MW}.

There is, however, no general agreement on the definition of entanglement
for systems with more than two parts.
Klyachko has proposed  \cite{Kly,KS} 
to regard as entangled the states which are {\em semi-stable}
for the action of the group of invertible local
filtering operations, also called SLOCC\footnote{for Stochastic Local
Operations
assisted with Classical Communication.},
\begin{equation}
G= SL(V_1)\times\cdots\times SL(V_k)
\end{equation}
in the sense of geometric invariant theory, which means those
states on which at
least one non trivial $G$-invariant polynomial does not vanish.
The point of introducing {\em geometric} invariant theory is that
this theory provides methods for characterizing such states without
explicitly computing the invariants. In order to explore the significance
of this property, the invariants have been explicited in the simplest cases
(up to 4 qubits, and 3 qutrits, with partial results for 5 qubits,
\cite{LT,LT4,VDdMV}).

One would also like to {\em quantify} entanglement. 
The non-locality properties of an entangled state
does not change under unitary operations acting independently on
each of its sub-systems. 
The idea of describing entanglement by means of local unitary
invariants is explored in \cite{Gra}, see also \cite{SM1,SM2}. 
However, except for the simplest systems, there are far too many orbits and
a complete classification is out of reach.

An intermediate possibility is to look at the $G$-orbits.
The knowledge of the $G$-invariant polynomials is not sufficient
to separate the $G$-orbits, and in general, one has to look
for the {\em covariants} in the sense of classical invariant theory.

In  \cite{BLT}, the algebra of   $G$-covariants
for $4$ qubits has been investigated, and a complete set
of generators has been obtained.

In the present article, we will explain how these results can be applied
to the calculation of bases of unitary invariants.
As an
application, we compute  bases of the spaces of 
local unitary and special unitary invariants  of degree $4$ of $k$ qubits for arbitrary $k$,
and recover the results of \cite{Gra2} for 3 and 4 qubits.

The paper is organized as follows. In Section \ref{SLOCCtoLUT} we
recall some background on SLOCC covariants, and we describe a method
allowing to obtain from them local unitary (LUT) and special
unitary (LSUT) invariants. Section \ref{simplest}
is devoted to the computation of the simplest LUT and LSUT
invariants from SLOCC covariants. Finally, we give some examples
and applications in Section \ref{examples}.

%%%%%%%%%%%%%%%%%%%%%%%%%%%%%%%%%%%%%%%%%%%%%%
%%%%
\section{Invariants and covariants of qubit systems}
\subsection{Group actions on state spaces}
Let $V=\C^2$ be the local Hilbert space of a two-state system
(a qubit), and
$\H=V^{\otimes k}$ be the state space of a system of $k$ qubits.
We shall regard it as the natural representation of the
group $G=G_{\SLOCC}=SL(2,\C)^k$,
known in  quantum information
theory  as the group of reversible local filtering operations,
or stochastic local
quantum operations assisted by classical communication (SLOCC) \cite{BW,Dur}.
This is a semisimple complex Lie group, whose representation theory
follows immediately from that of $SL(2,\C)$.
The maximal compact subgroup of $G$
is $K=G_{\LSUT}=SU(2)^k$, the group of local special unitary
transformations. We shall also be interested in arbitrary local
unitary transformations, which form the group $U=G_{\LUT}=U(2)^k$.

If $|j\>$,
$j=0,1$ is a basis of $V$, a state $|\Psi\>$ can be written as
\begin{equation}
|\Psi\>=\sum_{i_1,\ldots,i_k=0}^1 a_{i_1i_2\cdots i_k} |i_1i_2\cdots i_k\>
\end{equation}
where, as customary in the physics literature, 
\begin{equation}
|i_1i_2\cdots i_k\> = |i_1\>\otimes\cdots\otimes |i_k\>\,.
\end{equation}
It will be convenient to interpret such a state as a multilinear
form 
\begin{equation}
f(\xx)=f(x^{(1)},\ldots,x^{(k)})=\sum_{i_1,\ldots,i_k=0}^1 a_{i_1i_2\cdots i_k}
x^{(1)}_{i_1}\cdots x^{(k)}_{i_k}\,,
\end{equation}
where $x^{(j)}=(x^{(j)}_0,x^{(j)}_1)$ are pairs of variables.

The action of a $k$-tuple of matrices $\gg=(g^{(1)},\ldots,g^{(k)})$
on the various vector spaces introduced so far
is defined by $\gg\xx=\xx'$, $x'^{(i)}=g^{(i)}x^{(i)}$ and the components
$a'_{i_1i_2\cdots i_k}$ of $f'=\gg f$ are defined by the condition
\begin{equation}
\sum_{i_1,\ldots,i_k} a_{i_1i_2\cdots i_k}
x^{(1)}_{i_1}\cdots x^{(k)}_{i_k}
=
\sum_{i_1,\ldots,i_k} a'_{i_1i_2\cdots i_k}
x'^{(1)}_{i_1}\cdots x'^{(k)}_{i_k}\,.
\end{equation}

In the following, we shall be  interested in LUT and LSUT
invariants of a state $|\psi\>$, i.e. polynomial functions
$I(\aa,\bar\aa)$ in the components of $|\psi\>$, such that
\begin{equation}
I(\aa,\bar\aa)=I(\aa',\bar\aa')\,
\end{equation}
where $a'_{i_1\cdots i_k}$ are the components of $f'=\gg f$
for $\gg$ a LUT or a LSUT. Our main point will be the application
of the SLOCC invariant theory to the calculation of such unitary
invariants.

%%%%%%%%%%%%%%

\subsection{SLOCC Invariants}

The SLOCC invariants are the holomorphic polynomials $I(\aa)$
such that $I(\aa)=I(\aa')$ for $\gg\in G_\SLOCC$. Of course,
the squared modulus $|I|^2$ of a SLOCC invariant is a LSUT invariant,
but only a small subset of unitary invariant are of this form.

The methods of classical invariant theory can be applied to
the determination of the SLOCC invariants of $k$
qubits for small $k$. An important preliminary step is the
determination of the {\em Hilbert series}
\begin{equation}
h(t)=\sum_{d\ge 0}t^d \dim S^d(\H)^{G_\SLOCC},
\end{equation}
the generating series of the dimension of the space of homogeneous
polynomial
invariants of degree $d$.

For $k=3$, the only fundamental polynomial invariant of three qubits is 
known since the nineteenth century, it is the Cayley hyperdeterminant
\cite{LeP}, see also \cite{Mi}:
\begin{multline}
\Det(A)=(a_{000}^2a_{111}^2+a_{001}^2a_{110}^2+a_{010}^2a_{101}^2+a_{011}^2a_{100}^2)\\
-2(a_{000}a_{001}a_{110}a_{111}+a_{000}a_{010}a_{101}a_{111}+a_{000}a_{011}a_{1000}a_{1111}\\
+
a_{001}a_{010}a_{101}a_{110}+a_{001}a_{011}a_{110}a_{100}+a_{010}a_{011}a_{101}a_{100}\\
+
+4(a_{000}a_{011}a_{101}a_{110}+a_{001}a_{010}a_{100}a_{111})
\end{multline}

The polynomial invariants of four qubits were constructed
in \cite{LT}. Here, the Hilbert series is
\begin{equation}
h(t)=\frac{1}{(1-t^2)(1-t^4)^2(1-t)^6}\,.
\end{equation}

For five qubits, the Hilbert series and a few fundamental
invariants were  obtained in \cite{LT4}.

\subsection{Covariants}
%%%%%%%%%%%%

To construct the invariants, as well as for the more
difficult problem of classifying the orbits, one needs
of the classical notion of a covariant. 
A covariant $\Phi$ of $f$ is a
multi- homogeneous $G_\SLOCC$-invariant polynomial
in the form coefficients $a_{i_1\ldots i_k}$ and in the original
variables $x^{(i)}_j$, that is, an invariant in some space
\begin{equation}
\Phi\in S^{(d)}(\H)\otimes S^{\alpha_1}(V^*)\otimes\cdots\otimes S^{\alpha_k}(V^*)\,,
\end{equation}
where $\alpha$ is the multidegree of $\Phi$ in the $x^{(i)}_j$.

Clearly, a covariant can be interpreted as an equivariant map 
$u_\Phi$ from
the irreducible representation 
\begin{equation}
S_\alpha(V):=S^{\alpha_1}(V)\otimes\cdots\otimes S^{\alpha_k}(V)
\end{equation}
of $G$ to $S^d(\H)$. Such a map is uniquely determined by the image
of the highest weight vector $v_\alpha$ of $S_\alpha(V)$. This highest
weight vector is the coefficient of the highest monomial in $\Phi$,
classically called the {\em source} of the covariant. The coefficients
of the other monomials form a basis of weight vectors in the image
of $u_\Phi$.

The covariants form an algebra, which is naturally graded
with respect to $d$ and $\alpha$.
We denote by  ${\cal C}_{d;\alpha}$ the corresponding graded
pieces.
The knowledge of their dimensions $c_{d;\alpha}$
is equivalent to the decomposition of the character
of $S^d(\H)$ into irreducible characters of $G$,
and the knowledge of a basis of ${\cal C}_{d;\alpha}$
allows one to write down a Clebsch-Gordan series
with respect to $G$
for any polynomial in  $\aa$.
Also, it is known that the
equations of any $G$-invariant closed subvariety
of the projective space $\PP(\H)$ are given by the
simultaneous vanishing of the coefficients of some
covariants.

A modern introduction to classical invariant theory
can be found in the book \cite{Olv}.

\section{LUT-invariants from SLOCC-covariants\label{SLOCCtoLUT}}

\subsection{General construction}
A generating set of the algebra of the
polynomial covariants for the action of the SLOCC group can be in
principle computed by  a slight adaptation of the classical method
(the Cayley Omega process, see, e.g., \cite{Olv}). 
The covariants can be obtained recursively from the simplest one 
(the ground form $f$)
\begin{equation}
f=\sum_{i_1\cdots i_k}a_{i_1\cdots i_k}x^{(1)}_{i_1}\cdots
x^{(k)}_{i_k},
\end{equation}
by iterating an operation called transvection,  defined by
\begin{equation}\begin{array}{ll}
(\Psi,\Phi)^{\epsilon_1\cdots\epsilon_k}=&\tr\Omega_{x^{(1)}}^{\epsilon_1}
\cdots\Omega_{x^{(k)}}^{\epsilon_k}\Psi({x'}^{(1)},\cdots,{x'}^{(k)})\times\\&\times
\Phi({x''}^{(1)},\cdots,{x''}^{(k)})\end{array}
\end{equation}
where
\begin{equation}
\Omega_x=\det\left|\begin{array}{cc}\partial\over\partial {x'}_0&\partial\over\partial {x'}_1\\
\partial\over\partial {x''}_0&\partial\over\partial {x''}_1\end{array}\right|
\end{equation}
and $\tr: x', x''\rightarrow x$.

In practice, obtaining a
description of the algebra in terms of generators and syzygies
seems to be out of reach for more than
four qubits  \cite{BLT,LT4}.
Nevertheless it may be always
possible to compute the smallest covariants with relevant
geometric properties.
%%%%%%%%%%%%%%%%%%%%%%%%%

As already mentioned, a basis ${\rm Cov}_{k}$ 
of the space of the polynomial SLOCC-covariants  can be
identified with a basis of highest weight vectors in the symmetric algebra
$S(\H)$, so that one can write 
\begin{equation}\nonumber
S(V^{\otimes k})=\bigoplus_{\phi\in {\rm Cov}_{k}}V_\phi,
\end{equation}
where $V_\phi$ denotes the irreducible representation of $G$ whose
highest weight vector corresponds to the covariant $\phi$.

Polynomial invariants under LUT (resp. LSUT) live  in
$S(V^{\otimes k})\otimes S(V^{*\otimes k})$ and hence
 in
\begin{eqnarray}
\displaystyle\bigoplus_{\phi,\phi'\in{\rm Cov}_{k}\atop {\rm
deg}\phi={\rm deg}\phi'} \left(V_\phi\otimes
V_{\phi'}^*\right)^{\LU}\\\displaystyle (\mbox{resp.
}\bigoplus_{\phi,\phi'\in{\rm Cov}_{k}}\left(V_\phi\otimes
V_{\phi'}^*\right)^{\LSU})\,,
\end{eqnarray}
where {\rm deg}$\phi$ denotes the degree of $\phi$ in the
variables $a_{i_1\dots i_{k}}$. 

Note that if $\phi$ is a covariant whose
multidegree in the auxiliary variables is $(n_1,\dots,n_{k})$
the corresponding irreducible representation is
\begin{equation}
V_\phi\simeq S^{n_1}\left(\C^2\right)\otimes\cdots\otimes
S^{n_{k}}\left(\C^2\right).
\end{equation}
If $\phi$ and $\phi'$ are two polynomial covariants whose
respective multidegrees are $(n_1,\dots,n_{k})$ and
$(m_1,\dots,m_{k})$, then $V_\phi\otimes V_{\phi'}^*$ 
contains
polynomial invariants under LUT (resp. LSUT) if and only if
$n_1=m_1, \cdots, n_{k}=m_{k}$.
Moreover, combining the previous abstract nonsense identifying
covariants with $G$-highest weight vectors and $G$-equivariant maps, and
the canonical antilinear isomorphism of a Hilbert space with its hermitian
dual implies the following result:
 
\begin{proposition}\label{CompLUInv}
Denoting by $\Phi_{d,i}^\alpha$ a basis of SLOCC
covariants of degree $d$ in the entries of the tensor and
multidegree $\alpha$ in the auxiliary variables, one has:
\begin{enumerate}
\item The scalar products $\langle \Phi_{d,i}^\alpha|{{\Phi}_{d,j}^\alpha}\rangle$
with respect to the auxiliary variables, the $a_{i_1\ldots i_k}$ being regarded
as scalars, 
form a basis of the space of LUT invariants.
\item Similarly, the scalar products $\langle
\Phi_{d,i}^\alpha|{{\Phi}_{d',i}^\alpha}\rangle$ (where $d'$ is not necessarily
equal
to $d$), form a basis of the space of LSUT invariants.
\end{enumerate}
The hermitian scalar product induced by the one of $V$ can be calculated
by the formula
\begin{equation}
\nonumber \langle x_1\cdots x_m|y_1\cdots y_m\rangle={\rm
perm}\left(\langle x_i|y_j\rangle\right)
\end{equation}
if $\langle x_i|y_j\rangle=1$ when $x_i=y_j$ and $0$ otherwise.
\end{proposition}

This property should be of interest for the study of entanglement measures,
which are special LSUT invariants. Indeed, expressing such a measure
as a simple combination of scalar products of covariants with known geometric
properties might lead to interesting insights.
 
In the sequel, we will denote by $\Cov_{\SLOCC}(k)$
(resp. $\Inv_{\LU}$, $\Inv_{\LSU}$) the algebra of polynomial
SLOCC-covariants  (resp. LUT-invariants, LSUT-invariants). 
Note that these algebras are multigraded.
The space of multihomogeneous SLOCC-covariants (resp. LUT-invariants,  LSUT
-invariants) 
of degree $n$ in the $a_{i_1\cdots i_k}$ and ${\bf
 d}=(d_1,\cdots,d_k)$ in the auxiliary variables (resp. degree $n$ in the
 $a_{i_1\cdots i_k}$'s and the $\overline a_{i_1\cdots i_k}$'s, 
 degree $n_1$ in  the
 $a_{i_1\cdots i_k}$'s and  degree $n_2$ the
 $\overline a_{i_1\cdots i_k}$'s) will be denoted by $\Cov_{\SLOCC}(k;n;{\bf
 d})$ (resp. $\Inv_{\LUT}(k;n)$, $\Inv_{\LSUT}(k;(n_1,n_2))$).

\subsection{\label{Hilb}Hilbert series}

{F}rom Proposition \ref{CompLUInv}, we see that
the knowledge of the Hilbert
series of the SLOCC-convariants  allows one to compute the Hilbert
series of the LUT and LSUT-invariants.
We will denote  by
\begin{equation}
{h}_{\SLOCC}(k;z;{\bf u})=\sum \dim{\rm
Cov}_{\SLOCC}(n;k;{\bf d})z^n{\bf u}^{\bf d}
\end{equation}
where ${\bf u}^{\bf d}=u_1^{d_1}\cdots u_k^{d_k}$, the Hilbert
series of ${\rm Inv}_{\SLOCC}$. The Hilbert series of the algebras
${\rm Inv}_{\LU}$ and ${\rm Inv}_{\LSU}$ are obtained from 
${h}_{\SLOCC}(k;z;{\bf u})$ by the formulae
\begin{equation}
\begin{array}{rcl}{h}_{\LU}(k;z)&=&\sum_n\dim\Inv_{\LU}(k;2n)z^{2n}\\
&=&\left.{h}_{\SLOCC}(k;z^2;{\bf u})\odot{h}_{\SLOCC}(k;z^2;{\bf
u})\right|_{u_i=1}\\ 
&=& {\rm CT}_{z,u_1,\dots,u_k} 
\left\{{h}_{\SLOCC}(k;zt;(u_1,\dots,u_k))
{h}_{\SLOCC}(k;{z\over t};(u_1^{-1},\dots, u_k^{-1}) \right\}\,,
\end{array}
\end{equation}
where $\odot$ denotes the Hadamard product of the power series ring
$\C[[z,u_1,\dots,u_k]]$ 
(that is, 
$\uu^\alpha\odot\uu^\beta=\delta_{\alpha\beta}\uu^\alpha$), 
and ${\rm CT }_{x_1,\dots,x_n}f$ means the
constant term of the series $f$ with respect to
the variables $x_1,\dots, x_n$.

Similarly, one has
\begin{equation}
\begin{array}{rcl} {h}_{\LSU}(k;z)&=&\sum_{n_1,n_2}\dim\Inv_{\LUT}(k;(n_1,n_2))z^{n_1}\overline
z^{n_2}\\&=&\left.{h}_{\SLOCC}(k;z;{\bf
u})\odot_{\C[[z,\overline z]]} {h}_{\SLOCC}(k;\overline
z;{\bf u})\right|_{u_i=1}\\&=&{\rm CT}_{u_1,\dots,u_k}
\left\{{h}_{\SLOCC}(k;z;(u_1,\cdots,u_k){h}_{\SLOCC}(k;\overline z;(u_1^{-1},\cdots,u_k^{-1})\right\}\,,
\end{array}\end{equation}
where $\odot_{\C[[z,\overline z]]}$ denotes the Hadamard product
in $\C[[z,\overline z]][[u_1,\cdots,u_k]]$ ({\it i.e.}
considering $\C[[z,\overline z]]$ as the ring of scalars). 

Hence,
\begin{equation}
\dim{\rm Inv}_{\LU}(k;2n)=\sum_{\bf d}(\dim{\rm
Cov}_{\SLOCC}(n;k;{\bf d}))^2
\end{equation}
and
\begin{equation}
\dim{\rm Inv}_{\LSU}(k;(n_1,n_2))=\sum_{{\bf
d}}\dim{\rm Cov}_{\SLOCC}(n_1;k;{\bf d}) \dim{\rm
Cov}_{\SLOCC}(n_2;k;{\bf d}).
\end{equation}
Classical methods of invariant theory 
allows one to express the Hilbert series of algebras of covariants as a
constant term (see \cite{BLT} for an example). Hence, the Hilbert
series of unitary and special unitary invariants are
\begin{equation}\label{HLU}
{h}_{\LUT}(k;z)=\frac1{2^k}{\rm CT}_{t,\bf u}
\left\{{\displaystyle\prod_i(1-u_i^{-2})^2\over
\displaystyle\prod_{\alpha\in\{-1,+1\}^{k+1}\atop a=\pm 1
}(1-t^az\prod_i{\bf u}^\alpha) }\right\}
\end{equation}
and \begin{equation}\label{HLSU} {h}_{\LSUT}(k;z)=\frac1
{2^k}{\rm CT}_{\bf
u}\left\{\displaystyle\prod_i(1-u_i^{-2})^2\over
\displaystyle\prod_{\alpha\in\{-1,+1\}^k}\left[ (1-z{\bf
u}^\alpha) (1-\overline z{\bf u}^\alpha)\right]\right\}.
\end{equation}
These expressions have  been first derived by Beth et al. (unpublished,
see \cite{Gra2}), using a different method.

\section{Simplest invariants\label{simplest}}
\subsection{Dimension formulas for SLOCC-covariants\label{dim}}
The characters of the
irreducible polynomial representations of the group $GL(2,\C)^k$
are the products
\begin{equation}
s_{\underline\lambda}:=s_{\lambda^{(1)}}\cdots s_{\lambda^{(k)}}
\end{equation}
where $\underline\lambda=(\lambda^{(1)},\dots,\lambda^{(k)})$ is a
tuple of partitions $\lambda^{(i)}$ of length at most 2,
and $s_{\lambda^{(i)}}$ the corresponding irreducible character
of $GL(2,\C)$  i.e., a Schur function \cite{Macd}.
In particular, the
characters of the one-dimensional representations
\begin{equation}
\det{}^{\bf l}(\gg)= (\det g^{(1)})^{l_1} (\det g^{(2)})^{l_2}\cdots  (\det
g^{(k)})^{l_k}\,,
\end{equation}
containing the SLOCC invariants,
are the products
\begin{equation}
s_{(l_1l_1)}s_{(l_2l_2)}\cdots s_{(l_kl_k)}
\end{equation}
and the character of $GL(V)$  in $S^d(V)$ is $s_d$. 

Hence, the
dimension of the space of invariants of degree $d$ and weight ${\bf l}$,
which is also the multiplicity of the one-dimensional character
$\det^{\bf l}$ in $S^d(V)$, is given by the scalar product
\begin{eqnarray}
\dim {\rm Inv}_\SLOCC(d,k;{\bf l}) 
=
\< s_d\,|\, s_{(l_1l_1)} \cdots s_{(l_kl_k)}\>
\end{eqnarray}
of SLOCC characters  (where ${\bf l}=(l_1,\dots,l_k)$).
 We see that ${\rm
Inv}_\SLOCC(d;k;{\bf l})$ can be nonzero only if the condition
\begin{equation}
d=2l_1=2l_2=\cdots = 2l_k
\end{equation}
is satisfied. Hence,
\begin{equation}
\dim {\rm Inv}_\SLOCC(2l;k) =\sum_{\lambda\vdash
2l\atop l(\lambda)\le 2}
\frac1{z_\lambda}\chi_{\lambda}^{ll}\cdots\chi_{\lambda}^{ll}\,,
\end{equation}
where $\chi_{\lambda}^{ll}$ denotes the value of the irreducible
character $\chi^{ll}$ (labelled
by the partition $(l,l)$) of the symmetric group $\SG_{2l}$ on the
conjugacy class $\lambda=(1^{m_1}2^{m_2}\cdots n^{m_n})$,
and $z_\lambda= \prod_i i^{m_i}m_i!$, cf. \cite{Macd}.

In the same way, the  SLOCC-covariants of a $k$-qubit system
form the algebra
\begin{equation}
{\rm Cov}=[{S}({V}^{\otimes k})\otimes 
{S}(V^*\oplus\cdots\oplus V^* )]^{\SLOCC}
\end{equation}
which can be graded according both to the degree in the 
$a_{i_1\ldots i_k}$
and the multidegree in the auxiliary variables. A
similar reasoning  gives the dimension of the space of  covariants
of degree $d$ as
\begin{equation}\label{DimCov}
\dim{\rm Cov}_\SLOCC(d;k)=\sum_{\mu \vdash
n}\frac1{z_\mu}\left(\sum_{\lambda\vdash d\atop l(\lambda)\leq
2}\chi_\mu^\lambda\right)^k.
\end{equation}

Although impratical for finding closed forms of the Hilbert series,
these expressions are useful for computing the first terms.

\subsection{\label{cov123}Simplest SLOCC-covariants} 

The space of covariants of degree $1$ is generated by the {\em ground form}
 \begin{equation}\label{ground}
f=\sum_{0\leq i_1,\cdots,i_k\leq1}a_{i_1\cdots
i_k}x^{(1)}_{i_1}\cdots x^{(k)}_{i_k}.
\end{equation}
The dimension of the space of covariants of degree 2 of a $k$-
qubit system follows from formula (\ref{DimCov})
\begin{equation}
\dim{\rm Cov}_\SLOCC(2,k)=2^{k-1}.
\end{equation}
Observe that the only multihomogeneous covariants  in
this space have a multidegree in the auxiliary variables belonging
to $\{0,2\}^{k}$. If $\bf d$ is any tuple, we will denote by $|{\bf
d}|_a$ the number of occurrence of $a$ in $\bf d$. 
The dimension
of the space of covariants of degree 
${\bf d}=(d_1,\cdots,d_k)\in\{0,2\}^{k}$ 
in the auxiliary variables is
\begin{equation}
\dim {\rm Cov}_\SLOCC(2,k;{\bf d})
=
\langle (\chi^2)^{(k-|{\bf
d}|_0)}(\chi^{11})^{|{\bf
d}|_0}|\chi^2\rangle=\left\{\begin{array}{ll}0& \mbox{if }|{\bf
d}|_0\mbox{ is odd}\\ 1& \mbox{if }|{\bf d}|_0\mbox{ is
even}
\end{array}
\right..
\end{equation}
Hence, we can state the following result:
\begin{proposition}\label{covdim2}
The space of the covariants of degree $2$ of a $k$-qubit system
has dimension $2^{k-1}$ and is spanned by $f^2$ and the polynomials 
\begin{equation}
B_{\bf d}=(f,f)^{{2-d_1\over 2},\dots,{2-d_k\over 2}}\,,
\end{equation} 
where 
${\bf d}=(d_1,\cdots,d_k)\in\{0,2\}^{k}$ 
and 
$|{\bf d}|_0$ is even.
\end{proposition}

Note that if $k$ is odd, there are no invariants of degree $2$ and if
$k$ is even the invariants are all proportional to the
hyperdeterminant $(f,f)^{(1^k)}$.

The dimension of the space of covariants of degree 3
is, from formula (\ref{DimCov}),
\begin{equation}
\begin{array}{ll}
\dim{\rm Cov}_\SLOCC(3,k)&=\frac123^{k-1}+\frac12.
\end{array}
\end{equation}
The only covariants of degree $3$ in the entries of the tensor
have a multidegree ${\bf d}=(d_1,\dots,d_k)\in\{1,3\}^k$ in the
auxiliary variables. Let ${\bf d}=(d_1,\dots,d_k)\in\{1,3\}^k$ be
a multidegree, the dimension of the space of the covariants having
multidegree $\bf d$ is
\begin{equation}\label{dimcov3}\begin{array}{ll}
\dim{\rm Cov}_\SLOCC(3,k;{\bf d})&=\langle (\chi^{21})^{|{\bf
d}|_1}(\chi^3)^{|{\bf
d}|_3}|(\chi^3)\rangle\\&=\frac13\left(2^{|{\bf
d}|_1-1}+(-1)^{|{\bf d}|_1}\right)\end{array}
\end{equation}
if $|{\bf d}|_1>0$, and
\begin{equation}
\dim{\rm Cov}_\SLOCC(3,k;(3^k))=\langle
(\chi^3)^{k}|\chi^3\rangle=1.
\end{equation}
This implies that all the homogeneous covariants of multidegree
$(3^n)$ are proportional to $f^3$. From (\ref{dimcov3}), the
dimension of the space of the $k$-linear covariants of degree $3$
is 
\begin{equation}\label{dimcov11111}\begin{array}{ll}
\dim{\rm Cov}_\SLOCC(3,k;{\bf
d})&=\frac13\left(2^{k-1}+(-1)^{k}\right)\end{array}
\end{equation}

Let us denote by
$\{C_i\}_{i=1,\dots,\frac13(2^{k-1}+(-1)^{k-1})}$ a basis of
the space of  covariants of multidegree $(1^k)$. Applying
transvections with the ground form to the $C_i$, we obtain
invariants of degree $4$. Remark that the dimension of the space
of invariants of degree $4$ is equal to the dimension of the space
of multilinear covariants of degree $3$,
so that we recover a result of \cite{Bry}.  
We will denote by $(D_i)$
a basis of the space of SLOCC invariants of degree $4$.

\subsection{\label{inv4}Polynomial LUT-invariants of degree $4$}

{F}rom Proposition \ref{covdim2}, one can construct a basis
of the space of LUT invariants of degree $4$.
If ${\bf d}=(d_1,\dots,d_k)$, the dimension of 
${\rm Cov}_\SLOCC(2,k;{\bf d})$ is $0$ or $1$. 
Hence, the only possibilities are the squared norms
\begin{equation}\label{LUdim4}
{\bf B}_{\bf d}:=\langle B_{\bf d}|B_{\bf d}\rangle
\end{equation}
where ${\bf d}=(d_1,\cdots,d_k)\in\{0,2\}^{k}$ and $|{\bf d}|_0$
is even.

The dimension of the space is the coefficient of $z^4$ in
the Hilbert series
\begin{equation}
\dim{\rm Inv_{\LU}}(4,k)=\sum_{{\bf d}} \left( \dim {\rm
Cov}_\SLOCC(2,k;{\bf d})\right)^2=2^{k-1}.
\end{equation}
Thus, the following result holds:
\begin{proposition}
The space of the LUT invariants of degree $4$ of a  $k$-qubit
system has  dimension $2^{k-1}$ and is spanned by the polynomials
${\bf B}_{\bf d}$  and $\langle f|f\rangle^2$.
\end{proposition}

Furthermore, one can show that 
\begin{equation}
{\bf B}_{22\dots 2}=\langle f^2|f^2\rangle=2^k\langle
f|f\rangle^2-\sum_{{\bf d}\neq(2,\dots,2)}{\bf B}_{\bf d}\,.
\end{equation} 
Indeed, $\<f|f\>^2=\<f\otimes f|f\otimes f\>$, and the classical
Clebsch-Gordan series allows one to express $f\otimes f$ in terms
of transvectants, whence the result.

If we consider only the LSUT group, one has generators of bidegree
$(4,0)$, $(3,1)$, $(2,2)$, $(1,3)$, $(0,4)$ in the 
components of the state
and their conjugates. The subspace of bidegree $(2,2)$ is
the space of LUT-invariants of degree $4$:
\begin{equation}
{\rm Inv_{\LSU}}((2,2),k)={\rm Inv_{\LU}}(4,k).
\end{equation}
The subspace of bidegree $(4,0)$ if the space of the SLOCC
invariants of degree $4$
\begin{equation}
{\rm Inv_{\LSU}}((4,0),k)={\rm Inv_{\SLOCC}}(4,k).
\end{equation}
In the same way, the subspace of bidegree $(0,4)$ is the space of
the conjugates of the SLOCC invariants of degree $4$. A basis of
the space of the LSUT-invariants of bidegree $(3,1)$ can be obtained
from the scalar products $\langle C_i|f\rangle$.
\begin{proposition}
The subspace of LSUT-invariants of degree $4$ of a $k$-qubit system
has dimension \[\frac732^{k-1}-\frac43(-1)^{k-1}\] and is spanned
by the polynomials $D_i$ (bidegree $(4,0)$), ${\bf C}_i=\langle
C_i| f\rangle$ (bidegree $(3,1)$), ${\bf B}_{\bf d}$ (bidegree
$(2,2)$), $\overline{\bf C}_i$ (bidegree $(1,3)$) and $\overline
D_i$ (bidegree $(0,4)$).
\end{proposition}

\section{Examples \label{examples}}

\subsection{LUT-invariants and linear entropies\label{entrop}}

Let  $|\Psi\rangle=\sum a_{i_1\dots i_k}|i_1\cdots i_k\rangle$
be  a pure $k$-qubit state. 
Meyer and Wallach  \cite{MW} have
defined an  entanglement measure $\cal Q$ by
\begin{equation}
{\cal Q}(|\Psi\rangle)=\frac1k\sum_{i=1}^kD_1^{(i)}(|\Psi\rangle)
\end{equation}
where
%%%%%%%%%% INCOMPREHENSIBLE %%%%%%%%%%%
\begin{equation}D_1^{(i)}(|\Psi\rangle)=2\sum_{(\epsilon_1,\cdots,\epsilon_{k-1})\neq
(\epsilon'_1,\cdots,\epsilon'_{k-1})}\left|\left|
\begin{array}{cc}
\left(\epsilon_i^0|\Psi\right)&
\left({\epsilon'}_i^0|\Psi\right)\\
\left(\epsilon_i^1|\Psi\right)& \left({\epsilon'}_i^1|\Psi\right)
\end{array}\right|
\right|^2\,.\nonumber
\end{equation}
In this expression, 
$\left(\epsilon_i^\delta|\Psi\right)$ denotes the coefficient of
$\left|\epsilon_1\cdots\epsilon_{i-1}\delta\epsilon_{i+1}\cdots\epsilon_k\right\rangle$
in $|\Psi\rangle$, and the double bars  mean the squared modulus
of the determinant. 

%%%%%%%%%%%%%%%%%%%%%%%%%%%%%%%%%%%%%%%%
The interest of this measure resides in its
physical interpretation, related to the average purity of the
constituent qubits \cite{Br1} or the linearised form of the Von
Neumann entropy of a single qubit with the rest of the system.

Emary has remarked  \cite{Em1} that the functions $D_1^{(i)}$
are entanglement monotones, thus in particular  LU-invariants. 
Hence, each
$D_1^{(i)}$ can be written in terms  of squares of transvectants.
We have
\begin{equation}\label{D1}
D_1^{(i)}=\frac1{2^{k-2}}\sum_{d=(d_1,\cdots,d_k)\in\{0,2\}^k\atop
d_i=0} {\bf B}_d.
\end{equation}
Indeed, $D_1^{(i)}=\|\Phi_i\otimes \Phi_i\|^2$, where 
$\Phi_i=B_{22\cdots 202\cdots 2}$, and the result follows again
from the Clebsch-Gordan series.

Hence, in terms of our basis, the quantity ${\cal
Q}(|\Psi\rangle)$   has the simple expression
\begin{equation}
{\cal Q}(|\Psi\rangle)=\frac1{2^{k-2}k}\sum_{{\bf
d}=(d_1,\cdots,d_k)\in\{0,2\}^k}|{\bf d}|_0 {\bf B}_d.
\end{equation}

\subsection{LUT-invariants for  $3$-qubits\label{Grassl}} 

The algebra of covariants of  $3$ qubits is generated
by the polynomials \cite{LeP}
\begin{equation}
f:=\sum a_{i_1i_2i_3}x_{i_1}y_{i_2}z_{i3}\nonumber\end{equation}
\begin{eqnarray}
\nonumber H_x:=\left|\begin{array}{cc}{\partial^2 f\over\partial
y_0\partial z_0}& {\partial^2 f\over\partial y_1\partial z_0}\\
{\partial^2 f\over\partial y_0\partial z_1}& {\partial^2
f\over\partial y_1\partial z_1}\end{array}\right|\\
 \nonumber H_y:=\left|\begin{array}{cc}{\partial^2 f\over\partial x_0\partial
z_0}& {\partial^2 f\over\partial x_1\partial z_0}\\ {\partial^2
f\over\partial x_0\partial z_1}& {\partial^2 f\over\partial
x_1\partial z_1}\end{array}\right|\\
 \nonumber H_z:=\left|\begin{array}{cc}{\partial^2 f\over\partial x_0\partial
y_0}& {\partial^2 f\over\partial x_1\partial y_0}\\ {\partial^2
f\over\partial x_0\partial y_1}& {\partial^2 f\over\partial
x_1\partial y_1}\end{array}\right|
\end{eqnarray}
\begin{eqnarray}
\nonumber T:=\left|\begin{array}{cc}{\partial f\over\partial
x_0}&{\partial f\over\partial x_1}\\ {\partial H_x\over\partial
x_0}&{\partial H_x\over\partial x_1}
\end{array}\right|\end{eqnarray}\begin{eqnarray}
\nonumber\Delta:=(T,f)^{111}
\end{eqnarray}
{F}rom these polynomials, we can construct the following LU-invariants
\begin{eqnarray}
\nonumber {\bf A}_{111}:=\langle f|f\rangle\\ \nonumber {\bf
B}_{200}:=\langle
 \nonumber H_x|H_x\rangle\\
 \nonumber  {\bf B}_{020}:=\langle H_y|H_y\rangle\\
 \nonumber {\bf B}_{002}:=\langle H_z|H_z\rangle\\
 \nonumber  {\bf C}_{111}:=\langle T|T\rangle\\
 \nonumber {\bf D}_{000}:=\langle \Delta|\Delta\rangle\\
 \nonumber  {\bf F}_{222}:=\langle \Delta
 f^2|{T}^2\rangle
\end{eqnarray}
Grassl et al. \cite{Gra2} have computed a minimal system of seven
generators (denoted by $f_i$) of the algebra of LU invariants.
We shall give their expressions in terms of scalar products
of covariants.
 
The generator of degree $2$, $f_1$ is
clearly ${\bf A}_{111}$. To define generators of degree $4$ and
$6$, the authors  introduce the notation
%\begin{widetext}
\begin{equation}
f_{\sigma,\tau,\rho}:=\sum_{{{\bf i}=
(i_1,i_2,\dots,i_n),\atop {\bf j}=( j_1,j_2,\dots,j_n),}\atop {\bf
k}=(k_1,k_2, \dots,k_n)}{\bf a}_{\bf ijk}{\bf \overline a}_{{\bf
i}^{\sigma}{\bf j}^\tau{\bf k}^\rho}
\end{equation}
where ${\bf i}^\sigma=(i_{\sigma(1)},\dots,i_{\sigma(n)})$ and
${\bf a}_{\bf ijk}=a_{i_1j_1k_1}\cdots a_{i_nj_nk_n}$.
Their generators in degree $4$ and $6$ are
\begin{eqnarray}
 \nonumber f_2:=f_{(12),(12),{\rm Id}}={\bf A}_{111}^2-{\bf B}_{200}-{\bf
B}_{020}\\ \nonumber f_3:=f_{(12),{\rm Id},(12)}={\bf
A}_{111}^2-{\bf B}_{200}-{\bf B}_{002}\\ \nonumber f_4:=f_{{\rm
Id},(12),(12)}={\bf A}_{111}^2-{\bf B}_{020}-{\bf B}_{002}\\
 \nonumber f_5:=f_{(12),(23),(13)}={\bf A}_{111}^3+\frac32{\bf
C}_{111}-\frac32{\bf A}_{111}({\bf B}_{200}\\+{\bf B}_{020}+{\bf
B}_{002})\nonumber
\end{eqnarray}
Note that these invariants appear in many places in the literature, 
for example in \cite{Kem}.

The generator of degree $8$ is ${\bf D}_{000}$ and the generator
of degree $12$ is
\begin{eqnarray}
f_7:=&\overline\Delta\left([11,00]\{00,00\}-[11,00]\{11,11\}\right.\nonumber\\&
+[11,01]\{00,01\}+[11,10]\{00,10\}\nonumber\\
&+2[11,10]\{01,11\}-2[01,00]\{10,00\}\nonumber\\
&-[01,00]\{11,01\}-[10,00]\{11,10\}\nonumber\\
&-[10,01]\{00,00\}-[10,01]\{01,01\}\nonumber\\&+[10,01]\{10,10\}\left.+[10,01]\{11,11\}\right)^2\nonumber
\end{eqnarray}
where
$[i_1i_2,j_1j_2]=a_{i_1i_20}a_{j_1j_21}-a_{i_1i_21}a_{j_1j_20}$
and
$\{i_1i_2,j_1j_2\}=a_{i_1i_20}\overline{a_{j_1j_21}}+a_{i_1i_21}\overline{a_{j_1j_20}}$.
With our notations, one has
\begin{eqnarray}
f_7=\frac12{\bf D}_{000}\left(\frac32\left({\bf B}_{200}+{\bf
B}_{020}+{\bf B}_{002}\right)-{\bf A}_{111}^2\right)\nonumber\\
+2{\bf C}_{111}^2-4{\bf B}_{200}{\bf B}_{020}{\bf B}_{002}
+\frac18{\bf F}_{222}.\nonumber
\end{eqnarray}

The authors of \cite{Gra2} obtained the Hilbert series using residue
calculations in Magma.
We have been able to reproduce their results 
evaluating (\ref{HLU}) using a very
efficient algorithm due to Guoce Xin \cite{Xin} in a Maple
implementation. 
Summarizing, we have:
\begin{proposition}
The algebra of local unitary invariants 
pure $3$-qubit states is generated by ${\bf
A}_{111}$, ${\bf B}_{200}$, ${\bf B}_{020}$, ${\bf B}_{002}$,
${\bf C}_{111}$, ${\bf D}_{000}$ and ${\bf F}_{222}$. Its Hilbert
series is
\begin{equation}{h}_\LU(3;z)={1-t^{24}\over (1-t^2)(1-t^4)^3(1-t^6)(1-t^8)(1-t^{12})}\end{equation} which
where the numerator reflects the existence of a unique syzygy in degree $24$.
\end{proposition}

\subsection{Classification of the orbits under SLOCC
transformations \label{onion}} 

The normal forms of  $3$-qubit states
under SLOCC transformations
are known since 1881 \cite{LeP}. As shown on table \ref{class}, 
the SLOCC orbits can be characterized by the vanishing or non vanishing
of a set of four LU-invariants.
\begin{table}[t]
$$
\begin{array}{ccccc}
\hline &{\bf B}_{200}&{\bf B}_{020}&{\bf B}_{002}&{\bf D}_{000}\\
\hline |GHZ\rangle&\times&\times&\times&\times\\
|W\rangle&\times&\times&\times&0\\
|B_1\rangle=|001\rangle+|010\rangle&\times&0&0&0\\
|B_2\rangle=|001\rangle+|100\rangle&0&\times&0&0\\
|B_3\rangle=|010\rangle+|100\rangle&0&0&\times&0\\
|000\rangle&0&0&0&0\\\hline
\end{array}\nonumber
$$ \caption{SLOCC orbits of three qubit states.\label{class}}
\end{table}
In the table, a
$\times$ means the non-nullity of the invariant. Hence, 
(\ref{D1}) implies that in this case,
the ``onion classification'' \cite{Mi} can be 
described only in terms of proper entanglement measures (entanglement monotones),
see fig \ref{onion}.

\begin{figure}[h]
\begin{center}\resizebox{5cm}{5cm}{\includegraphics{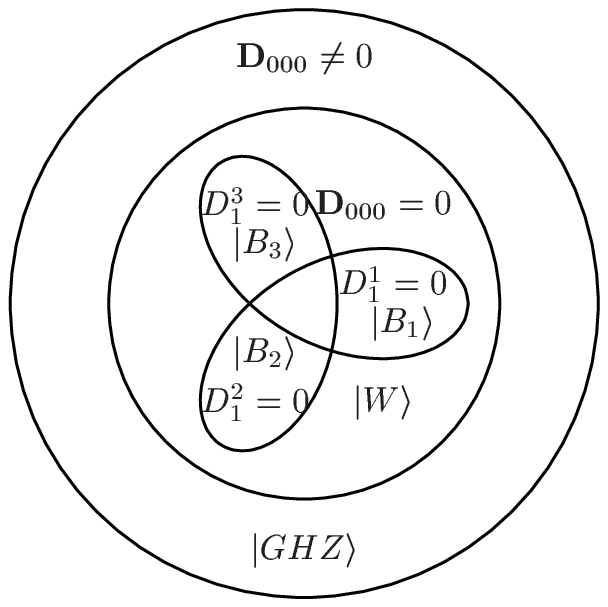}}\end{center}
\caption{SLOCC orbit structure for $3$-qubits \label{onion}}
\end{figure}

\subsection{LSUT-invariants for  $3$-qubits} 

Another unpublished result of Grassl et al.  \cite{Gra2}
can be recovered from (\ref{HLSU}) by means of Xin's algorithm \cite{Xin}.
It is the Hilbert series of the algebra of LSUT-invariants of three qubits,
\begin{equation}
{z^5\overline{z}^5+z^3\overline{z}^3+z^2\overline{z}^2+1\over
(1-z\overline{z})(1-z^4)(1-z^2\overline{z}^2)^2(1-\overline{z}^4)(1-z\overline{z^3})(1-z^3\overline{z})}
\end{equation}
This expression suggests that the algebra has a Cohen-Macaulay structure
with $6$ primary invariants of respective bidegrees
($(1,1),(0,4),(2,2),(2,2),(4,0),(1,3)$ and $(3,1)$ and $3$ secondary
invariants of bidegree $(2,2), (3,3), (5,5)$. The set of primary
invariants is 
$${\cal P}=\{{\bf A}_{111}, f_2, f_3, \Delta,
\overline{\Delta}, s_2:=\langle A,T\rangle, \overline{s_2}\}\,.$$
Computing the Jacobian of 
${\bf A}_{111},f_2, f_3, \Delta, 
\overline{\Delta}, s_2,\overline{s_2},a_{000},\dots,a_{111},
\overline{a_{000}}$ 
with the random numerical values given in table \ref{eval},
one finds 
$$-53279560564736-243669580382208i\neq 0.$$ 
This implies that the polynomials 
${\bf A}_{111} ,f_2, f_3, \Delta, \overline{\Delta}, s_2$
and 
$\overline{s_2}$ are algebraically independent.
\begin{table}
$$\begin{array}{|c|c|c|c|c|c|c|c|}\hline
a_{000}&a_{001}&a_{010}&a_{011}&a_{100}&a_{101}&a_{110}&a_{111}\\
\hline 3+3i& 3+3i&3+3i&2+i&3+2i&1+2i&2+3i&3+i\\
\hline\end{array}$$ \caption{\label{eval}  Random values of the
$a_{ijk}$'s.}
\end{table}
The set of secondary invariants is ${\cal S}=\{f_4,f_5,f_4f_5\}$.
The polynomial $f_4$ (resp. $f_5$) is linearly independent of all
algebraic combinations of bidegree $(2,2)$ (resp. $(3,3)$) of
elements of $\cal P$. Furthermore, one has two syzygies involving
respectively $f_4^2$ and $f_5^2$:
\begin{equation}
8f_1f_5-6f_4f_2+3f_4^2-3|\Delta|^2+3f_2^2-6f_4f_3+f_1^4+3f_3^2-6f_3f_2-12|s_2|^2
=0 
\end{equation}
and
\begin{multline}
-18f_4f_1^4-18f_3f_1^4-18f_2f_1^4+11f_1^6+18\overline{\Delta}s_2^2-36|s_2|^2f_3+18\Delta\overline{s_    2}^2
-72f_4f_3f_2\\
+30f_4f_2f_1^2+30f_4f_3f_1^2-36|s_2|^2f_2+60|s_2|^2f_1^2+3f_4^2f_1^2+3f_3^2f_1^2+30f_3f_2f_1^2
-36|s_2|^2f_4\\
+3f_2^2f_1^2-3|\Delta|^2f_1^2+16f_5^2=0\,.
\end{multline}
This implies the following property:
\begin{proposition}
The algebra of LSUT invariants of three qubits
is a free module over a polynomial
algebra (Cohen-Macaulay structure)
\begin{equation}
\Inv_{\LSUT}=\displaystyle\bigoplus_{{\bf c}\in {\cal S}}
\C[{\cal P}]{\bf c}
\end{equation}
\end{proposition}

\subsection{LUT invariants of four qubits}

\begin{table}[t]
$$\begin{array}{|c|c||c|c||c|c||c|c||c|c||c|c||c|c|} \hline
i&a_i&i&a_i&i&a_i&i&a_i&i&a_i&i&a_i&i&a_i\\\hline
 2&0&4&0&6&6&8&46&10&110&12&344&14&844\\
 16&2154&18&4606&20&9397&22&16848&24&28747&26&44580&28&65366\\
 30&88036&32&111909&34&131368&36&145676&38&149860& 40&145676&42&131368\\
 44&111909&46&88036&48&65366& 50&44580&52&28747&54&16848&56&9397\\
 58&4606&60&2154&62&844&64&344&66&110&68&46&70&6\\
 72&0&74&0&76&1&&&&&&&& \\\hline\end{array}$$
 \caption{\label{LU4} Hilbert series of LUT invariants for $4$ qubits: values of the $a_{i}$.}
\end{table}
\begin{table}[t] $$
\begin{array}{|c|c||c|c||c|c||c|c||c|c|}
\hline
(i,j)&a_{i,j}&(i,j)&a_{i,j}&(i,j)&a_{i,j}&(i,j)&a_{i,j}&(i,j)&a_{i,j}\\
\hline (0,0)&1&(1,3)&-1&(2,2)&2&(2,4)&6&(2,6)&9\\
 (2,8)&4&(2,10)&3&(3,3)&7&(3,5)&12& (3,7)&12\\
 (3,9)&7&(3,11)&2&(3,13)&-3&(4,4)&28&(4,6)&42\\
 (4,8)&52&
 (4,10)&36&(4,12)&12&(4,16)&1&(5,5)&43\\
 (5,7)&79&
 (5,9)&92&(5,11)&36&(5,13)&-1&(5,15)&-12\\(5,17)&-6&
 (5,19)&-1&(6,6)&132&(6,8)&199&(6,10)&161\\(6,12)&53&
 (6,14
 )&-9&(6,16)&-27&(6,18)&-10&(7,7)&214\\(7,9)&236&
 (7,11)&129&(7,13)&-12&(7,15)&-83&(7,17)&-63\\
 (7,19)&-15&(7,21)&-2&(8,8)&339&(8,10)&289& (8,12)&110\\
 (8,14)&-115&(8,16)&-169&(8,18)&-82&(8,20)&-21&(8,22)&-3\\
 (9,9)&306&(9,11)&160&(9,13)&-154&(9,15)&-363&(9,17)&-253\\
 (9,19)&-82&(9,21)&-12&(9,23)&3&(10,10)&268&(10,12)&-96\\
 (10,14)&-513&(10,16)&-510&(10,18)&-234&(10,20)&-37&(10,22)&12\\
 (10,24)&3&(11,11)&-126&(11,13)&-676&(11,15)&-818&(11,17)&-465\\
 (11,19)&-85&(11,21)&76&(11,23)&41&(11,25)&4&(12,12)&-681\\
 (12,14)&-1045&(12,16)&-763&(12,18)&-221&(12,20)&133&(12,22)&154\\
 (12,24)&36&(12,26)&3&(13,13)&-1152&(13,15)&-985&(13,17)&-359\\
 (13,19)&265&(13,21)&424&(13,23)&216&(13,25)&41&(13,27)&3\\
 (14,14)&-1094&(14,16)&-543&(14,18)&245&(14,20)&705&(14,22)&496\\
 (14,24)&154&(14,26)&12&(14,28)&-3&(15,15)&-569&(15,17)&318\\
 (15,19)&1058&(15,21)&992&(15,23)&424&(15,25)&76&(15,27)&-12\\
 (15,29)&-2&(16,16)&233&(16,18)&1188&(16,20)&1334&(16,22)&705\\
 (16,24)&133&(16,26)&-37&(16,28)&-21&(17,17)&1333&(17,19)&1734\\
 (17,21)&1058&(17,23)&265&(17,25)&-85&(17,27)&-82&(17,29)&-15\\
 (17,31)&-1&(18,18)&1736&(18,20)&1188&(18,22)&245&(18,24)&-221\\
 (18,26)&-234&(18,28)&-82&(18,30)&-10&(19,19)&1333&(19,21)&318\\
 (19,23)&-359&(19,25)&-465&(19,27)&-253&(19,29)&-63&(19,31)&-6\\
 (20,20)&233&(20,22)&-543&(20,24)&-763&(20,26)&-510&(20,28)&-169\\
 (20,30)&-27&(20,32)&1&(21,21)&-569&(21,23)&-985&(21,25)&-818\\
 (21,27)&-363&(21,29)&-83&(21,31)&-12&(22,22)&-1094&(22,24)&-1045\\
 (22,26)&-513&(22,28)&-115&(22,30)&-9&(23,23)&-1152&(23,25)&-676\\
 (23,27)&-154&(23,29)&-12&(23,31)&-1&(23,33)&-3&(24,24)&-681\\
 (24,26)&-96&(24,28)&110&(24,30)&53&(24,32)&12&(25,25)&-126\\
 (25,27)&160&(25,29)&129&(25,31)&36&(25,33)&2&(26,26)&268\\
 (26,28)&289&(26,30)&161&(26,32)&36&(26,34)&3&(27,27)&306\\
 (27,29)&236&(27,31)&92&(27,33)&7&(28,28)&339&(28,30)&199\\
 (28,32)&52&(28,34)&4&(29,29)&214&(29,31)&79&(29,33)&12\\
 (30,30)&132&(30,32)&42&(30,34)&9&(31,31)&43&(31,33)&12\\
 (32,32)&28&(32,34)&6&(33,33)&7&(33,35)&-1&(34
,34)&2\\(36,36)&1&&&&&&&&\\ 
\hline
\end{array}
$$
\caption{\label{LSU4}Hilbert series of LSU invariants for
$4$-qubits: values of the $a_{ij}=a_{ji}$. }
\end{table} 
Again, we have computed
the Hilbert series of LUT covariants of $4$ qubits 
by means
of Xin's algorithm. 
This allowed us to reproduce another
result of \cite{Gra2}.
\begin{equation}
{h}_{\LUT}(4;z)={P(z)\over Q(z)}
\end{equation}
with $P(z)=1+\sum_{ij}a_{i}z^i\overline z^j$ where the $a_{i}$ are
given in Table \ref{LU4} and
$$Q(z)=(1-z^{10})(1-z^8)^4(1-z^6)^6(1-z^4)^7(1-z^2)\,.$$ 
This suggests that the algebra has a  Cohen-Macaulay structure
with $19$ primary invariants and $1449936$ secondary invariants.
The complete knowledge of the generators is with no doubt out of
reach, nevertheless one can  compute the first primary
invariants using the covariants obtained in a previous paper \cite{BLT}.
The simplest one is the scalar square of the ground form 
$${\bf A}_{1111}=\langle f|f \rangle\,.$$
There are 6 bi-quadratic linear covariants of degree 2 and 1
invariant. This allows to construct unitary invariants of degree
 $4$:
 \begin{eqnarray}
 \nonumber {\bf B}_{2200}=&\langle B_{2200}|B_{2200}\rangle,\\
 \nonumber {\bf B}_{2020}=&\langle B_{2020}|B_{2020}\rangle,\\
 \nonumber {\bf B}_{2002}=&\langle B_{2002}|B_{2002}\rangle,\\
  \nonumber {\bf B}_{0220}=&\langle B_{0220}|B_{0220}\rangle,\\
  \nonumber {\bf B}_{0202}=&\langle B_{0202}|B_{0202}\rangle,\\
  \nonumber {\bf B}_{0022}=&\langle B_{0022}|B_{0022}\rangle,\\
  \nonumber {\bf B}=&B_{0000}\overline{B_{0000}}.
 \end{eqnarray}
The polynomial 
$\langle f^2|f^2\rangle$ 
is algebraically dependent
of the other ones: 
$$\langle f^2|f^2\rangle=16{\bf A}^2-\left({\bf
B}_{2200}+{\bf B}_{2020}+{\bf B}_{2002}+{\bf B}_{0220}+{\bf
B}_{0220}+{\bf B}_{0202}+{\bf B}_{0022}+{\bf B}\right)\,.$$ 
The space of linear covariants of degree $3$ is spanned by two
quadrilinear polynomials
\begin{eqnarray}
 C^1_{1111}=(f,B_{2200})^{1100}\nonumber,\\
 C^2_{1111}=(f,B_{2020})^{1010}\nonumber
\end{eqnarray}
and four cubico-trilinear covariants \cite{BLT}
\begin{eqnarray}
 C_{3111}=(f,B_{2200})^{0100}\nonumber\\
 C_{1311}=(f,B_{2200})^{1000}\nonumber\\
 C_{1131}=(f,B_{2020})^{1000}\nonumber\\
 C_{1113}=(f,B_{2002})^{1000}.\nonumber
\end{eqnarray}
With these polynomials, one can construct a set of twenty
generators for the space of unitary invariants of degree $6$: 

\medskip
\noindent
${\bf A}^3$,\\ 
${\bf A}{\bf B}$, 
${\bf A}{\bf B}_{2200}$, 
${\bf A}{\bf B}_{2020}$, 
${\bf A}{\bf B}_{2002}$, 
${\bf A}{\bf B}_{0220}$,
${\bf A}{\bf B}_{0202}$, 
${\bf A}{\bf B}_{0022}$, \\
$\langle C^1_{1111},C^1_{1111}\rangle$, 
$\langle C^1_{1111}, C^2_{1111}\rangle$, 
$\langle C^1_{1111}, fB_{0000}\rangle$,
$\langle C^2_{1111},C^1_{1111}\rangle$, \\
$\langle C^2_{1111},C^2_{1111}\rangle$, 
$\langle C^2_{1111}, fB_{0000}\rangle$, 
$\langle fB_{0000}, C^1_{1111}\rangle$,
$\langle fB_{0000},C^2_{1111}\rangle$, \\
$\langle C_{3111},C_{3111}\rangle$, 
$\langle C_{1311}, C_{1311}\rangle$,
$\langle C_{1131}, C_{1131}\rangle$, 
$\langle C_{1113}, C_{1113}\rangle$. 

\medskip
The series suggests that the algebra has a
Cohen-Macaulay structure with $19$ primary invariants 
(one of degree $2$, seven of degree $4$, four of degree $8$ and one
of degree $10$). The polynomials
\begin{equation}\begin{array}{l}
{\bf A}_{1111},\nonumber\\
 {\bf B}, {\bf B}_{2200}, {\bf B}_{2020}, {\bf B}_{2002}, {\bf
 B}_{0220}, {\bf B}_{0202}, {\bf B}_{0022}\nonumber\\
 \langle C_{1111}^1,C_{1111}^1\rangle, \langle C_{1111}^1,AB\rangle,
 \langle C_{3111},C_{3111}\rangle,  \langle
 C_{1311},C_{1311}\rangle, \langle C_{1131},C_{1131}\rangle,  \langle C_{1113},C_{1113}\rangle
 \\
  \langle D_{4000},D_{4000}\rangle, \langle
  D_{0400},D_{0400}\rangle,\langle D_{0040},D_{0040}\rangle, \langle
  D_{0004},D_{0004}\rangle\nonumber\\
  \langle E_{3111},E_{3111}\rangle  \nonumber
\end{array}\end{equation}
where
\begin{eqnarray}
D_{4000}=(A,C_{3111})^{0111},\nonumber\\
D_{0400}=(A,C_{1311})^{1011},\nonumber\\
D_{0040}=(A,C_{1131})^{1101},\nonumber\\
D_{0004}=(A,C_{1113})^{1110},\nonumber\\
 D_{2200}=(A,C_{3111})^{1011} \nonumber\\
\mbox{and } E_{3111}=(A,D_{2200})^{1100}\nonumber
\end{eqnarray}
are algebraically independent and hence  good
candidates to be primary invariants.

\subsection{LSUT invariants of $4$ qubits}

Finally, we can compute the Hilbert series of LSUT invariants of $4$ qubits
by the same method, and again recover a  result of
\cite{Gra2}.
\begin{equation}
{h}_{\LSUT}(k;z,\overline z)={P(t)\over Q(t)}
\end{equation}
with $P(t)=\sum_{ij}a_{ij}z^i\overline z^j$, the $a_{ij}$ being
given in Table \ref{LSU4}
 and
\begin{equation}
\nonumber
\begin{array}{rl}
Q(t)=&(1-z\overline z)(1-z^2\overline z^2)^4(1-z^3\overline
z^3)(1-z^2)(1-z^4)^2(1-z^6)\\ &(1-\overline z^2)(1-\overline
z^4)^2(1-\overline z^6)(1-z^3\overline z)^3(1-z\overline
z^3)^3(1-z^2\overline z^4)\\ &(1-z^4\overline z^2)(1-z\overline
z^5)(1-z^5\overline z))
\end{array}
\end{equation}

\section{Conclusion}

We have proposed a new method to compute  bases of
the algebras of unitary invariants of qubit systems. 
This method involves as an intermediate step the calculation of the SLOCC
covariants, which have a more transparent geometrical meaning (at least
in small degrees), and leads naturally to new bases in which the known
entanglement measures tend to admit rather simple expressions.

The complete description of the algebra of unitary 
invariants for pure $k$-qubits is definitely
out of reach of any computer
system for $k>3$. This impossibility means that such
a study is not physically relevant and that only a few invariants 
with interesting geometrical properties will be significant in the
realm of quantum information theory. Finally, a natural question is
whether these constructions can be extended to  mixed states.

\end{document}